\documentclass[%
 reprint,
 amsmath,amssymb,
 aps,
pra,
]{revtex4-2}
\usepackage[
    colorlinks=true,
    linkcolor=blue,           % color of internal links (sections, figures, etc.)
    urlcolor=blue,            % color of URLs
    citecolor=blue
]{hyperref}
\usepackage{soul}

\renewcommand*{\phi}{\varphi}

\newcommand*{\ket}[1]{| #1 \rangle}

\newcommand{\rg}[1]{{\color{blue} #1 }}

\usepackage{graphicx}% Include figure files
\usepackage{dcolumn}% Align table columns on decimal point
\usepackage{bm}% bold math
\begin{document}

\preprint{APS/123-QED}

%\title{D2-line spectroscopy of $^{173}$Yb$^{+}$ ion using ultracold atoms of $^{6}$Li}
%\title{Precision spectroscopy of a trapped $^{173}$Yb$^+$ ion using a bath of ultracold atoms}
\title{Precision Spectroscopy of a Trapped Ion Using a Bath of Ultracold Atoms}

\author{Egor Kovlakov}%\affiliation{\affA}
%\email{ekovlakov@gmail.com}
\author{Rene Gerritsma}%\affiliation{\affA}\affiliation{\affB}%
\affiliation{Van der Waals-Zeeman Institute, Institute of Physics, University of Amsterdam, 1098 XH Amsterdam, Netherlands}

\date{\today}% It is always \today, today,

\begin{abstract}
We perform laser spectroscopy of a trapped ion that is not directly laser cooled. An ultracold atomic bath continuously cools the internal degrees of freedom of the ion to its hyperfine ground state via spin-exchange collisions while a cotrapped ${}^{174}$Yb$^+$ ion sympathetically cools the external degrees of freedom. Successful laser excitation is detected via state-selective charge transfer following a collision with an atom and subsequent ion loss. As a demonstration of this spectroscopic technique, we probe the $6^2S_{1/2}\rightarrow 6^2P_{3/2}$ transition at 329~nm in  $^{173}$Yb$^+$ and measure the magnetic and electric hyperfine interaction constants for the $6^2P_{3/2}$ state to be $A=-241(1)$~MHz and $B=1460(8)$~MHz, respectively. Our results are in agreement with a previous measurement obtained in a hollow-cathode discharge experiment but are a factor of 6–9 more precise. The techniques demonstrated in this work may be extended to perform precision spectroscopy on ions with complex level structures with applications in fundamental physics.
\end{abstract}
\maketitle

%\tableofcontents

%\section{\label{sec:level1}Introduction}

Neutral buffer gas cooling of trapped ions has been a widely used technique for a long time and finds applications in precision spectroscopy and quantum chemistry~\cite{Itano:1995,Gerlich:1995,Bhatt:2019}. While laser cooling generally outperforms neutral buffer gas cooling in terms of the low translational energies achieved, buffer gas cooling can offer unique advantages in particular situations. Most importantly, buffer gas cooling allows us to straightforwardly cool both the external and internal degrees of freedom of the trapped ions, while sympathetic cooling with laser-cooled atomic ions  only cools the former~\cite{Hudson:2009}. Moreover, systems combining laser-cooled trapped ions and ultracold atoms have been realized in different experimental setups in the last two decades~\cite{Zipkes:2010,Schmid:2010,Tomza:2019,Deiss:2024}. These systems have demonstrated buffer gas cooling capabilities that rival laser cooling for carefully chosen experimental parameters~\cite{Feldker:2020,Schmidt:2020}. Reaching collision energies in the 10~$\mu$K regime, the trapped ion-atom mixture can be used to resolve the spin dynamics of colliding atoms and ions. This allows quantum state preparation at the level of individual hyperfine and Zeeman states~\cite{Ratschbacher:2013,Sikorsky:2018,Fuerst:2018,Weckesser:2021,Katz:2022} with applications in precision spectroscopy and clocks~\cite{Tomza:2020}.

\begin{figure}%[b!]
%\label{Fig_sequence}
\centering{\includegraphics[width=\linewidth]{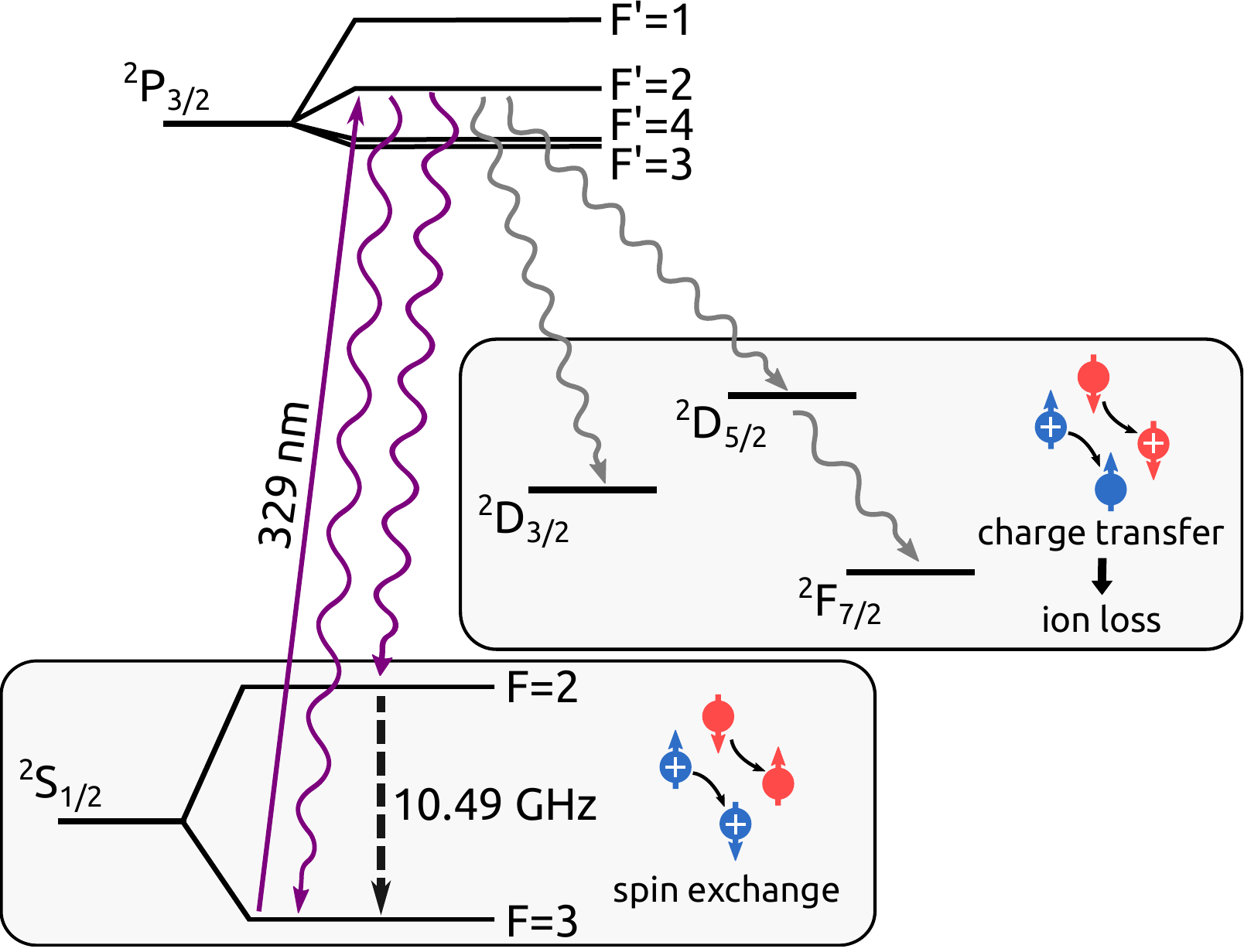}}
\caption{The level scheme of $^{173}$Yb$^{+}$ (not to scale). We probe the $^{2}S_{1/2} \rightarrow {}^{2}P_{3/2}$ transition using a sequence of 329~nm laser pulses, after which the ion has a considerable probability to spontaneously decay (wavy lines) into one of the metastable states $^2D_{3/2}$ or $^2D_{5/2}$ and subsequently end up in the extremely long-lived $^2F_{7/2}$ state. This leads to a highly probable charge transfer collision with a $^6$Li atom, resulting in ion loss. The time between pulses is set to allow the ion to be cooled back to the $F=3$ ground state via a spin-exchange collision with a $^6$Li atom. In this way, leakage out of the 329~nm cycle by spontaneous decay into the $F=2$ state is eliminated.}\label{fig:level_structure}
\end{figure}

Here, we demonstrate laser spectroscopy of a trapped ion that is immersed in a bath of ultracold $^6$Li atoms. Spin-exchange collisions between the ion and the atoms continuously prepare the ground hyperfine level of the spectroscopy ion, while we use state-selective charge transfer~\cite{Ratschbacher:2012,Haze:2015,Saito:2017,Joger:2017,Xing:2024} to detect laser excitation as a function of frequency. We demonstrate this spectroscopic technique through observation of the $^{2}S_{1/2} \rightarrow {}^{2}P_{3/2}$ transition in $^{173}$Yb$^+$, which has previously only been studied in a hollow-cathode discharge experiment~\cite{berends1992hyperfine}. The working principle of our experiment is sketched in Fig.~\ref{fig:level_structure}. While ${}^{173}$Yb$^+$ can also be directly probed with methods developed for single ions~\cite{Roman:2021,dellaert2025metastable,Jiang:2026,Yu:2026}, the demonstrated technique can be generalized to perform precision spectroscopy on ions with complex level schemes with applications in fundamental physics.   

\begin{figure}
\centering{\includegraphics[width=\linewidth]{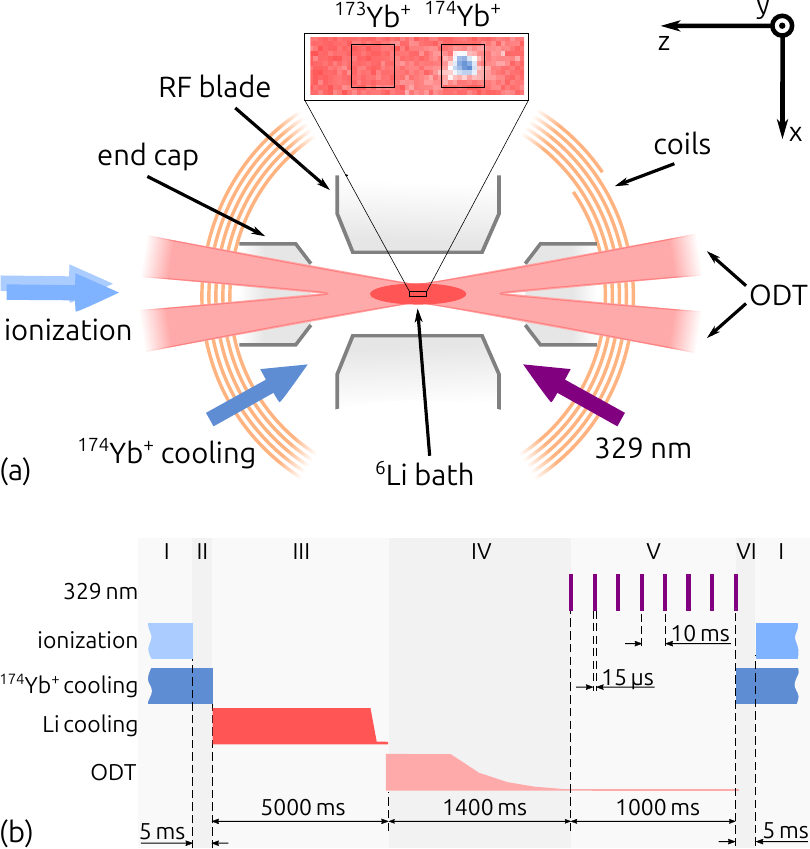}}
\caption{\rg{(a)~}Setup schematics. Laser beams are
focused into the center of the Paul trap, where the cloud of Li atoms is loaded into \rg{a crossed} ODT. The inset shows a fluorescence image of a two ion crystal composed of a dark ion ($^{173}$Yb$^+$ ) and a bright ion ($^{174}$Yb$^+$).~(b)~Experimental timing sequence diagram:~(I)~ions loading;~(II)~ions cooling;~(III)~loading of ODT with Li atoms;~(IV)~evaporation cooling of Li;~(V)~spectroscopy pulses;~(VI)~$^{174}$Yb$^+$ ion position detection.}\label{fig:setup}
\end{figure}

\emph{Setup}---The Yb$^+$-Li mixture is prepared in a hybrid ion-neutral trap as depicted in Fig.~\ref{fig:setup}(a) and described more extensively in Ref.~\cite{Hirzler:2020}. In order to load Yb$^+$ ions into the linear Paul trap, we employ isotope-selective two-step photoionization with lasers at 399~nm wavelength for the excitation of the $^{1}S_{0} \rightarrow {}^{1}P_{1}$ transition in neutral Yb and at 369~nm for a nonresonant excitation into the continuum. For the first step, we use two different 399~nm lasers detuned from each other by $\approx$~250 MHz to switch between $^{173}$Yb$^+$ and $^{174}$Yb$^+$ isotopes~\cite{Dipankar:2005}. For the second step, we apply a single 369~nm laser, which is also used for driving the $^{2}S_{1/2} \rightarrow {}^{2}P_{1/2}$ Doppler-cooling transition of the $^{174}$Yb$^+$ isotope.

Each experimental run begins with loading a single $^{174}$Yb$^+$ ion and Doppler cooling it to about 0.5~mK, the fluorescence from this "bright" ion is collected on an electron-multiplying charge-coupled device camera. Next, we open the shutter of the $^{173}$Yb$^+$ photoionization beam and wait for the $^{174}$Yb$^+$ ion to shift axially away from the trap center. This displacement signals the formation of a two-ion crystal with a sympathetically cooled "dark" $^{173}$Yb$^+$ ion. Then we block the $^{174}$Yb$^+$ cooling laser beams to prepare the bright ion in the $^{2}S_{1/2}$ ground state. After that, we prepare a cloud of fermionic $^{6}$Li atoms in a magneto-optical trap located about 20~mm below the ion trap. We optically pump the atoms into the lowest two magnetic sublevels $\ket{F=1/2, m_F= \pm 1/2}$, and load the resulting equal spin mixture into a 1070~nm optical dipole trap (ODT) formed by beams with a 40~µm waist, crossed $\approx$~200~µm below the ion. After forced evaporative cooling at a magnetic field of 780~G, close to the 832~G Feshbach resonance \cite{zurn2013precise}, we obtain about $2 \times 10^4$ $^{6}$Li atoms at a temperature of $\approx 5$~µK. Finally, we immerse both ions in the $^{6}$Li bath by repositioning the ODT using piezocontrolled mirrors. We let the atoms and ions interact for 1~s, while applying a train of laser pulses at 329~nm%, each with a width of 15 $\mu$s
. The experimental timing sequence is depicted in Fig.~\ref{fig:setup}~(b).

For 329~nm light generation, we use a frequency-quadrupled amplified diode laser. After the first doubling cavity, the 658~nm light is coupled into a high-bandwidth fiber electro-optic modulator (EOM) to produce sidebands in the 1–3~GHz range. These sidebands are used for an offset lock to an external ultrastable optical cavity via the Pound-Drever-Hall technique. A portion of the doubling cavity output is tapped off before the EOM and coupled into a wavelength meter for absolute frequency determination of the frequency-doubled light with a 10~MHz uncertainty, corresponding to a 20~MHz uncertainty for the frequency-quadrupled light. The wavelength meter is calibrated using a 689~nm laser from a neighboring laboratory that is locked to the narrow $^1S_0\rightarrow {}^3P_1$  transition in $^{88}$Sr~\cite{Courtillot:2005}.

\emph{Experiment}---We scan the frequency and shape the pulses of the 329~nm light using an acousto-optic modulator (AOM) in a double-pass configuration with a center frequency of 210~MHz and a bandwidth of 80~MHz. The signals for the AOM and the EOM are generated by a two-channel microwave generator, which is stabilized to a 10~MHz reference signal from a Rb clock. The AOM is calibrated to achieve constant optical power output during frequency scans. The 329~nm light is linearly polarized ($\sigma^+ + \sigma^-$) to drive $\Delta m_F = \pm 1$ transitions only, eliminating the first-order Zeeman shift by averaging transitions with opposite magnetic-field dependence. The quantization axis is defined by an applied magnetic field parallel to the $y$ axis, provided by a pair of low-inductance coils. We switch them on rapidly at the start of the atom-ion interaction to produce a constant homogeneous field of 2.6 G.

%The ion trap operates at a driving frequency $\Omega \approx 2\pi \times $ 1.85 MHz and trap frequencies $(\omega_x,\omega_y,\omega_z) \approx 2\pi \times (191, 196, 112)$ kHz, where z is the direction along the axis of the Paul trap.

For the 329~nm laser interrogation we used 100 laser pulses, each with a width of $\tau_\text{pulse}=15$~µs. The number of repetitions was chosen large enough to promote decay to the metastable states $^2D_{3/2}$ and $^2D_{5/2}$ via the branching fractions of the state ${}^2P_{3/2}$ as described below. %that were previously measured in $^{174}$Yb$^+$~\cite{Feldker:2018}. 
The pulse length was chosen such that $\tau_\text{pulse}\ll \Gamma_\text{L}^{-1}$, where $\Gamma_\text{L}\sim$~10$^2$~s$^{-1}$ denotes the Langevin collision rate between atoms and ions estimated from the measured atomic density~\cite{Feldker:2020}. The short pulses minimize atom-ion collisions during laser interrogation which would complicate and broaden the measured spectrum via molecular effects.   

The interval of 10~ms between the pulses was set to give the ion time to be cooled back to the $F=3$ ground state via a spin-exchange collision with a $^6$Li atom. This process closes leakage out of the 329~nm cycle due to spontaneous decay of the form $F'=3\rightarrow F=2$ and $F'=2\rightarrow F=2$, with $F'$ the total angular momentum in the ${}^2P_{3/2}$ state. Spin-exchange measurements of other Yb$^+$ isotopes immersed in $^6$Li reveal rates that approach the Langevin rate $\Gamma_\text{SE}\sim\Gamma_\text{L}$~\cite{Fuerst:2018,Feldker:2020}. We anticipate the small difference in reduced mass $\mu$ to lead to negligible deviations in scattering lengths between different Yb$^+$ isotopes~\cite{Tomza:2015,Feldker:2020}. For example, $\mu_{6,173}/\mu_{6,171}\sim$~1.000\,39 such that comparable spin exchange rates as in the combination $^{171}$Yb$^+$-$^6$Li may be expected.
Experimentally, we find setting the time in between pulses $>$~2~ms to be sufficient, in rough agreement with our estimated $\Gamma_\text{L}$.

Spin exchange is accompanied by energy release to the ions of magnitude $\Delta E_\text{SE}\sim h \Delta\nu_\text{hfs}\mu_{6,173}/m_\text{i}$, with $h$ Planck's constant, $\Delta \nu_\text{hfs} = 10\,491.7$~MHz the hyperfine splitting in the ${}^2S_{1/2}$ ground state~\cite{Munch:1987} and $m_\text{i}$ the  mass of the ion. Here, we neglected the absorption of energy due to the much smaller hyperfine splitting of the atoms. Distributed over the 6 degrees of freedom of the two-ion crystal, the energy release leads to a temperature increase of $\sim$~3~mK. Spin-exchange rates between atoms and ions are nearly independent of collision energy over a wide range~\cite{Sikorsky2018plb,Cote2018sot,Fuerst:2018,Feldker:2020}, such that this energy increase will not affect the outcome of subsequent exchanges.

The branching fractions into the states $^2D_{3/2}$ and $^2D_{5/2}$ from ${}^2P_{3/2}$ were measured in $^{174}$Yb$^+$ to be 0.17(1)~\% and 1.08(5)~\%, respectively~\cite{Feldker:2018}. The $^2D_{5/2}$ state  subsequently decays predominantly to the extremely long-lived ${}^2F_{7/2}$ state~\cite{Taylor:1997}. 
Collisions with atoms while the ion is in a metastable state have an increased chance to result in charge transfer. Using $^{174}$Yb$^+$ we have measured the charge transfer rate of the states $^2D_{3/2}$ and $^2F_{7/2}$ to be $\sim 0.03 \times \Gamma_\text{L}$ and $\sim 0.4 \times \Gamma_\text{L}$, respectively~\cite{Joger:2017}. In contrast, the charge transfer rate for the ${}^2S_{1/2}$ ground state is estimated to be negligible, $\lesssim 10^{-4}\times\Gamma_\text{L}$~\cite{Hirzler:2020}. Following charge transfer, the Li$^+$ ion is quickly lost from the ion trap, as its small mass makes stable trapping not possible. We observe the loss of the $^{173}$Yb$^+$ ion by subsequent imaging of the position of the $^{174}$Yb$^+$ probe ion with a shift to the center of the trap indicating ion loss.

\begin{figure}
\centering{\includegraphics[width=\linewidth]{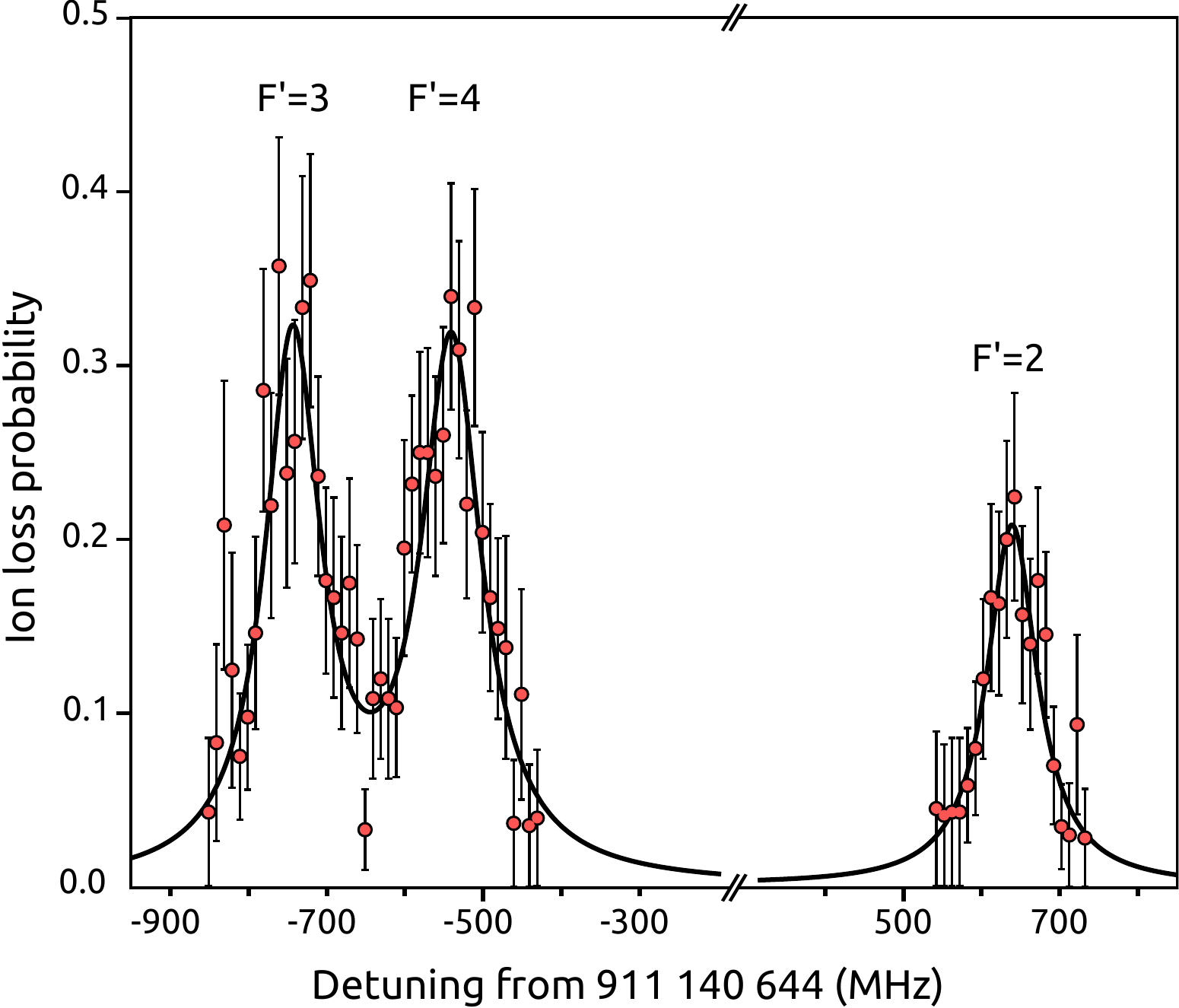}}
\caption{The ion-loss spectrum of $^2S_{1/2}\rightarrow {}^2P_{3/2}$ transition in $^{173}$Yb$^+$ from $F=3$ to $F'=\{2,3,4\}$. The data are fitted by a sum of Lorentzian functions. The experimental data points represent an average of at least 20 experimental realizations and error bars denote the standard error.}\label{fig:spectra}
\end{figure}

\emph{Results}---In Fig.~\ref{fig:spectra} we present the probability that the dark $^{173}$Yb$^+$ ion is lost after interaction with atoms as a function of the spectroscopy light frequency. We observe three peaks, corresponding to the three dipole-allowed hyperfine $^2S_{1/2}\rightarrow {}^2P_{3/2}$ transitions from $F=3$ to $F'=2$, $F'=3$ and $F'=4$.

The measured data are fitted using a sum of three Lorentzian functions with nine free parameters: the overall frequency offset, amplitudes, widths, and peak center positions. The latter were expressed in terms of the two hyperfine-structure constants $A$ and $B$ using the following expression for the hyperfine splitting relative to the line centers~\cite{Corney:1978}: 
\begin{equation}
    \Delta E = C_1 \times A + C_2 \times B,\\
\end{equation}
where 
\begin{equation}
    C_1 = \dfrac{F(F+1) - I(I+1) - J(J+1)}{2},\\
\end{equation}

\begin{equation}
    C_2 = \dfrac{3K(K+1)-4I(I+1)J(J+1)}{8I(2I-1)J(2J-1)},
\end{equation}
$A$ is the nuclear magnetic dipole hyperfine-structure constant, $B$ is the nuclear electric quadrupole hyperfine-structure constant, and $K$ is given by
\begin{equation}
    K = F(F+1) - I(I+1) - J(J+1).
\end{equation}

For the total angular momentum $J=3/2$ and $F'=\{2,3,4\}$ the corresponding $C_1 = \{-13/4,-1/4,15/4 \}$ and $C_2 = \{-2/20,-11/20,5/20 \}$.

\begin{table}[h]
    \caption{\label{tab:table1} Hyperfine constants for the $^{2}P_{3/2}$ state of $^{173}$Yb$^+$. The quoted errors in this work are statistical.}
        \begin{ruledtabular}
            \begin{tabular}{lcc}
            Reference & $A$ (MHz)& $B$ (MHz)\\
            \hline
            This work & $-241(1)$ & 1460(8)  \\
            Experiment\,\cite{berends1992hyperfine} & $-245(9)$ & 1460(50) \\
        Experiment\,\cite{Krebs:1956} & $-255(30)$ & 18(3) \\
        Theory\,\cite{Mani:2011} & $-88.973$ & 1839.779
        \\
         Theory\,\cite{Maartensson:1994} & $-107$ & 1780
        \\
            \end{tabular}
        \end{ruledtabular}
\end{table}

The constants $A$ and $B$ obtained from the fit are listed in Table~\ref{tab:table1}  alongside previously reported values. The uncertainty in the estimated constants correspond to the standard errors of the fit.

Each peak measurement takes approximately 6\,h and is performed on different days.  The drift of the reference cavity ($\sim 4$~kHz/day) is considered to be insignificant during this time. The difference in peak amplitudes could be explained by day-to-day fluctuations in atom-ion overlap and 329~nm laser power. Taking into account the ground-state hyperfine splitting $\Delta \nu_\text{hfs}$, we estimate the absolute frequency of the $^2S_{1/2}\rightarrow {}^2P_{3/2}$ central line to be $911.136\,272(20)$~THz.

\begin{figure}
\centering{\includegraphics[width=\linewidth]{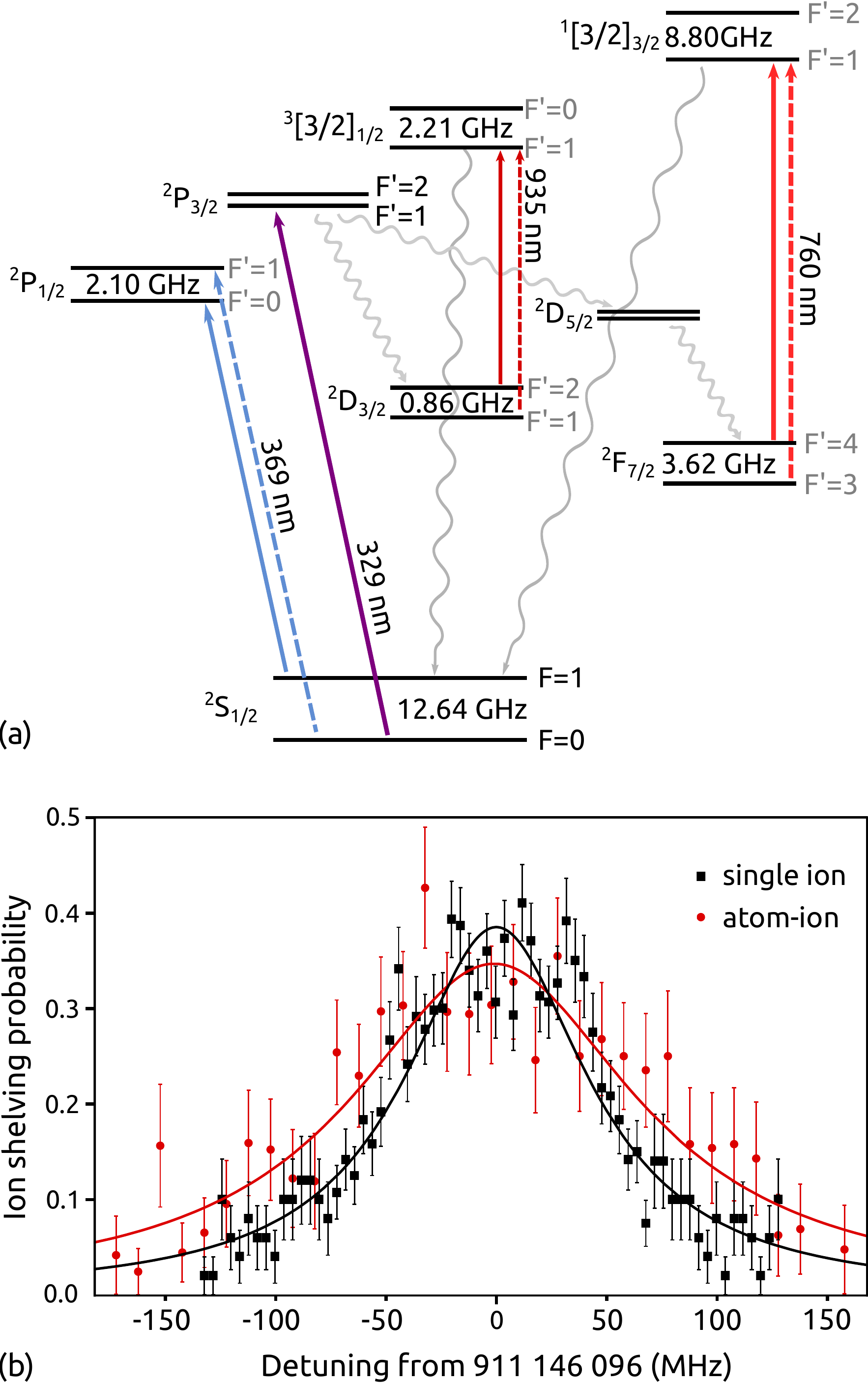}}
\caption{(a)~Energy levels and relevant transitions in $^{171}$Yb$^{+}$ (not to scale). Straight lines represent the processes induced by the lasers, and wavy lines those by spontaneous emission.~(b)~Frequency scan of the $^2S_{1/2}\rightarrow {}^2P_{3/2}$ transition in $^{171}$Yb$^+$ from $F=0$ to $F'=1$: with a single ion (black) and atoms (red). The data is fitted by Lorentzian functions. The experimental data points represent an average of at least 20 experimental realizations and error bars denote the standard error.}\label{fig:combined}
\end{figure}

Prior to measurements with the $^{173}$Yb$^+$ ion, we validated our method using the $^{171}$Yb$^+$ isotope and subsequently performed reference spectroscopy without atoms. For these ion-only measurements, a single  $^{171}$Yb$^+$ ion is trapped and Doppler cooled using lasers at 369 and 935~nm (repump). The hyperfine splittings of the $^{2}S_{1/2} \rightarrow {}^{2}P_{1/2}$ transition are bridged with resonant EOMs at 14.74 and 2.11~GHz. Similarly, the hyperfine splittings of the repump $^{2}D_{3/2} \rightarrow {}^{3}[3/2]_{1/2}$ transition are bridged with a resonant EOM at 3.07~GHz~\cite{Olmschenk:2007}, effectively closing the cooling cycle [see Fig.~\ref{fig:combined}(a)].

We apply 100 laser pulses at 329~nm with the same $\tau_\text{pulse}=15$~ µs, and monitor the ion fluorescence with a camera to detect shelving into the $^{2}F_{7/2}$ state. In contrast to the atom-assisted measurements, coupling between the $F=0$ and $F=1$ ground-state sublevels is provided by the 369~nm cooling light via optical pumping into $F=0$ state after each spectroscopy pulse, which is achieved by applying the 2.10~GHz EOM while keeping the 14.74~GHz EOM off. In the case of successful dark state detection, we pump the ion back into the ground state by exciting the $^{2}F_{7/2} \rightarrow {}^{1}[3/2]_{3/2}$ transition with a diode laser near the wavelength of 760~nm coupled into a fiber EOM. Applying the tones at 3.62~GHz, we are able to depopulate both hyperfine states $F=3$ and $F=4$ of the $^{2}F_{7/2}$ manifold.

The measured dark state population versus drive frequency of the AOM for the $^2S_{1/2}\rightarrow {}^2P_{3/2}$ transition in $^{171}$Yb$^+$ from $F=0$ to $F'=1$ is plotted in Fig.~\ref{fig:combined}~(b). It is accompanied by the results of the charge-exchange measurements with the two-ion crystal of $^{171}$Yb$^+$ and $^{174}$Yb$^+$ ions and Li atoms, where the $^{171}$Yb$^+$ ion plays the role of the dark ion of interest. We can conclude that the centers of both peaks coincide within the fit uncertainty. 

The resonance linewidth is significantly broader in the presence of atoms: $79\pm16$~MHz vs $50\pm5$~MHz. This broadening may be attributed to kinetic energy release following inelastic atom-ion collisions. In particular, up to $\sim$~100 spin-exchange collisions lead to $\sim$~20~MHz Doppler broadening. Additional Doppler broadening may occur due to quenching collisions. In cases where the ion was brought to a metastable state, and the collision with the atom does not lead to charge transfer, the path back to the ground state may release a lot of kinetic energy when it is not fully radiative. This has, for instance, been observed in the combination Rb-Sr$^+$~\cite{Ben-shlomi:2020}. Unfortunately, the complexity of the Yb$^+$ due to the open 4$f$ shell in the excited states precludes modeling of this effect~\cite{Joger:2017}. 

We also expect atomic density fluctuations to cause broadening. Since the current method relies on ion loss measurements, the atomic bath can be used only once such that each individual run takes  $>$7~s. Measuring the full dataset therefore spanned several days over which both atom number and atom-ion overlap may vary. We avoided apparent line center shifts by randomizing the frequency sampling within 80~MHz windows and then switching at random between these windows.  

Finally, all measured distributions are power broadened. The width of the resonance measured with a single ion corresponds to the saturation parameter of the spectroscopy light $s \approx 7.6$. In this estimate, we used the lifetime of of the ${}^{2}P_{3/2}$ level, $\tau = 7.23$~ns, as reported in~\cite{Biemont:1998}.

\emph{Conclusions}---Our results for the hyperfine constants are in agreement with the experimental results from Ref.~\cite{berends1992hyperfine} and are more precise. According to the authors of Ref.~\cite{berends1992hyperfine}, the 2 orders of magnitude difference in  $B$ from Ref.~\cite{Krebs:1956} is attributed to the rather limited resolution of the spectrum in that experiment. At the same time, the experimental results deviate from theoretical predictions in Ref.~\cite{Mani:2011} (relativistic coupled-cluster calculations) and Ref.~\cite{Maartensson:1994} (many-body perturbation theory calculations). The discrepancy may originate from the difficulty in modeling the strong mixing of the ${}^2P_{3/2}$ state with the 4$f^{13}$5d6s configuration and in particular the nearby $^3[3/2]_{3/2}$ state~\cite{Porsev:2012}. The magnitude of this effect may be important when considering hyperfine-induced electric dipole contributions to clock transitions involving the metastable states  $^2F_{7/2}$ and $^3[3/2]_{5/2}$~\cite{Dzuba:2016,Yu:2026,Ackerman:2026}.

Compared to quantum logic spectroscopy~\cite{Schmidt:2006}, our method as demonstrated falls short in terms of precision (see, for example ,~\cite{Gebert:2015}). However, we may borrow elements of quantum logic spectroscopy to make it competitive. The main improvement will be to abandon ion loss measurement in favor of measuring the released energy following a collision with an ion that was successfully excited to a metastable state. In our setup this would amount to measuring a $\sim$~3~mK temperature increase following a single spin-exchange collision~\cite{Katz:2022} and after scattering on the order of just one photon. This would eliminate Doppler broadening as a source of error. In this case the ultracold gas can be recycled, increasing the data rate and reducing fluctuations in atom number and ion-atom alignment. In such a setup the reactive Li can be replaced by a more deeply bound atom. The prime candidate for this is Cr as it has a large ionization energy~\cite{Tomza:2015}, has an electron spin in its ground state such that it can act as a spin bath~\cite{Ratschbacher:2013,Sikorsky:2018,Fuerst:2018}, it has a smaller mass than the ions of interest mentioned below~\cite{Cetina:2012,Tomza:2019} and it has been brought into the ultracold regime~\cite{Griesmaier:2005}. Finally, the atomic gas preparation stage can be separated from the science cell to reduce magnetic field fluctuations and improve the vacuum.

The reported spectroscopic technique is particularly useful for atomic ions with a dense fine- or hyperfine spectrum as the atoms can effectively prepare these ions in their ground state. Certain molecular ions may also be considered, provided chemical reactions with the ultracold buffer gas are sufficiency suppressed for the molecular ground state~\cite{Hall:2012,Tomza:2019,Hudson:2009,Deiglmayr:2012,Mohammadi:2021,Deiss:2024,Patel:2026}. 

Possible applications may be found in fundamental physics. For example, the system can be used to study isotope shifts and perform King plot analysis~\cite{Fuchs:2025} in species with a large number of isotopes without nuclear spin such as Er$^+$, Dy$^+$ and Gd$^+$. These ions have complex spectra featuring many low-lying metastable states owing to the open 4$f$ shell configurations. For such ions, the ultracold buffer gas may offer an attractive alternative to tedious laser operations to prepare the ground state and metastable state detection. Other ions of interest include Y$^+$ and Ac$^+$~\cite{Kawasaki:2026} for probing variations in the fine-structure constant and studies of heavy elements~\cite{Geehan:2026}. Moreover, the setup may be generalized to study the quantum dynamics of ultracold atoms interacting with ion species besides the commonly used earth alkaline and alkali like ions~\cite{Tomza:2019}. 

%\section{\label{sec:level1}Acknowledgments}
%\begin{acknowledgments}
\emph{Acknowledgments}---We thank the group of Florian Schreck for making available the Sr 689~nm laser for wavelength calibration. We thank Jook Walraven for comments on the manuscript, Micha{\l} Tomza for fruitful discussions and Nella Diepeveen for measuring the ultrastable cavity drift. This work was supported by the Dutch Research Council (DOI: 10.61686/DNTBM55710 343 and DOI: 10.61686/FUUKZ21999). The data and metadata presented in this work are available at~\cite{Dataset}.
%\end{acknowledgments}

\bibliographystyle{apsrev4-1}
\bibliography{main}% Produces the bibliography via BibTeX.

\end{document}